# Interactions of magnetized plasma flows in pulsed-power driven experiments

L G Suttle[1], G C Burdiak[1], C L Cheung[1], T Clayson[1], J W D Halliday[1], J D Hare[1], S Rusli[1], D Russell[1], E Tubman[1], A Ciardi[2], N F Loureiro[3], J Li[4], A Frank[5] and S V Lebedev[1]

[1] Blackett Laboratory, Imperial College London, SW7 2BW, United Kingdom
[2] Sorbonne Université, Observatoire de Paris, Université PSL, CNRS, LERMA, F-75005 Paris, France
[3] Plasma Science and Fusion Center, Massachusetts Institute of Technology, Cambridge, Massachusetts 02139, USA
[4] Northwest Institute of Nuclear Technology, Xi'an 710024, China
[5] Department of Physics and Astronomy, University of Rochester, Rochester, New York 14627, USA

Email: l.suttle10@imperial.ac.uk

**Abstract**

A supersonic flow of magnetized plasma is produced by the application of a 1 MA-peak, 500 ns current pulse to a cylindrical arrangement of parallel wires, known as an inverse wire array. The plasma flow is produced by the $\mathbf{J} \times \mathbf{B}$ acceleration of the ablated wire material, and a magnetic field of several Tesla is embedded at source by the driving current. This setup has been used for a variety of experiments investigating the interactions of magnetized plasma flows. In experiments designed to investigate magnetic reconnection, the collision of counter-streaming flows, carrying oppositely directed magnetic fields, leads to the formation of a reconnection layer in which we observe ions reaching temperatures much greater than predicted by classical heating mechanisms. The breakup of this layer under the plasmoid instability is dependent on the properties of the inflowing plasma, which can be controlled by the choice of the wire array material. In other experiments, magnetized shocks were formed by placing obstacles in the path of the magnetized plasma flow. The pile-up of magnetic flux in front of a conducting obstacle produces a magnetic precursor acting on upstream electrons at the distance of the ion inertial length. This precursor subsequently develops into a steep density transition via ion-electron fluid decoupling. Obstacles which possess a strong private magnetic field affect the upstream flow over a much greater distance, providing an extended bow shock structure. In the region surrounding the obstacle the magnetic pressure holds off the flow, forming a void of plasma material, analogous to the magnetopause around planetary bodies with self-generated magnetic fields.

## 1. Introduction

The interactions of supersonic, magnetized plasma flows can be challenging to predict, as they inherently combine the physics of shocks with the intricacies of two-fluid magnetohydrodynamic and magnetized kinetic effects. These effects can add preferential directions to processes such as thermal conduction, affect particle trajectories and plasma compressibility, perturb particle energy distributions, and ultimately dictate the global structure of the systems in which the interactions take place [1]. Magnetized plasma flows occur frequently in space and astrophysics, where for example super-Alfvénic outflows from stars and other energetic sources sweep extended magnetic field lines into the paths of various astrophysical objects and interstellar media [2]. Magnetized flows are also relevant in inertial confinement fusion (ICF) research, as embedded fields can be induced by strong gradients, and can affect system convergence [3–6]. Thus, applications such as these systems require an understanding of the fundamental processes involved in magnetized plasma flow interactions.

Creating magnetized high energy density plasma flows for laboratory studies of these processes can be difficult, as this often requires embedding a magnetic field into an initially unmagnetized plasma. When using an external magnetic field for this task, there must be sufficient time for the field to penetrate into and couple to the plasma before the intended interactions of the flow take place [7,8]. Experimental measurements need to verify that the flux is truly frozen to the plasma, to distinguish it from the case of interactions of unmagetized plasmas in the presence of an external magnetic field. It can also be challenging to ensure that the field penetrates uniformly, rather than seeding density, temperature and flux anisotropy during the interaction of the initially separate components [9–12].

In this paper we demonstrate the setup of a pulsed-power driven, inverse wire array as a means of creating a long-lasting flow of plasma, with a magnetic field which is strongly coupled to the flow from the plasma source. This is a versatile system as a super-sonic (and in many cases super-Alfvénic) plasma and flux travel together into an open region which is initially free from any background field. We can therefore use this flow to design interactions of varying geometries and obstacle types, which we provide several examples of in section 3. The first two of these examples (namely the study of magnetic reconnection in colliding flows [section 3.1] and interaction of magnetized flow with conducting obstacles [section 3.2]) are from established experiments performed with this platform and we aim to give only a brief summary of the main results from the works referenced. In the third example [section 3.3] we present the first results from a new set of experiments studying the interaction of the magnetized flow with strongly magnetized obstacles.

## 2. Production of a magnetized plasma flow

*2.1 The inverse wire array as a source of magnetized plasma*

In the experiments described in this paper a magnetized plasma flow is produced by the setup of an "inverse wire array"* [13]. This consists of a cylindrical cage of thin wires connected in parallel between two electrodes, with the negative electrode (cathode) extending along the central axis of the array as depicted in figure 1(a). The scale of the array is typically 15 – 20 mm in length by 16 – 20 mm in diameter, with a 5 – 8 mm diameter cathode. The wire array is fielded on the MAGPIE pulsed-power generator [14], which acts as a high-impedance current-source, providing the load (wire array) with a 500 ns duration current pulse, peaking at 1-1.4 MA. This current pulse resistively heats and ionizes the wires, forming an ablated plasma from the wire surfaces [15,16]. For a typical array of 16 wires, the diameter of the wires is selected in the range of 10 – 300 µm, depending on the wire material, which provides a continuous supply of plasma for the duration of the current pulse, without depleting the material in the inner core of the wire (which remains cold and dense).

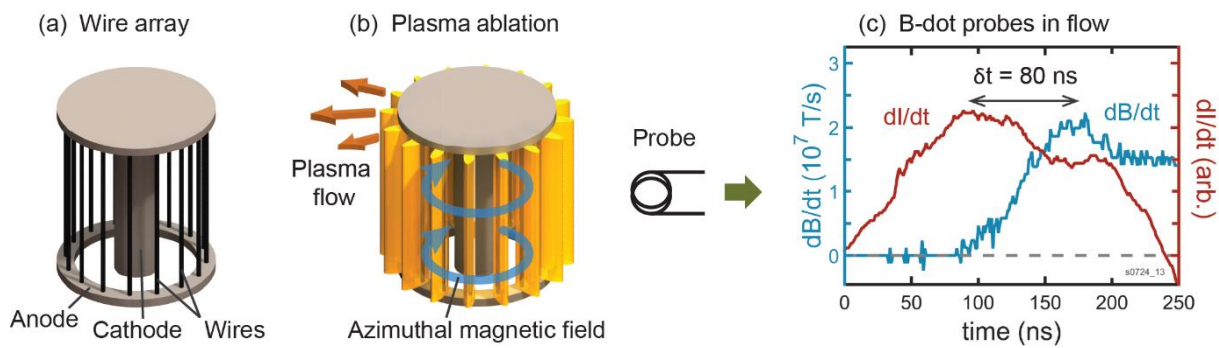

**Figure 1.** (a) 3D diagram of the inverse wire array, consisting of a cylindrical arrangement of thin conducting wires, connected between two electrodes. The cathode directs the return current along the axis of the array. (b) Ablation of the wire array, by the applied current pulse, creates a radially directed plasma flow with an azimuthal embedded magnetic field. (c) Inductive probes placed externally to the array detect a magnetic field signature which replicates the form of the driving current pulse and is offset by the time-of-flight of the plasma.

In addition to creating this plasma, the current also generates a strong magnetic field in the setup. Prior to the start of ablation (first 20 – 50 ns of the applied current pulse) the initial magnetic topology of this arrangement of current carrying conductors is dominated by a strong azimuthal field in the internal cavity surrounding the cathode, with smaller private magnetic fields surrounding the wires due to their fractional share of the current. Outside the array at radial positions greater than the inter-wire spacing this initial field distribution rapidly drops to <1% of the cavity field strength, as the field topology converges to that of a coaxial line [13,17]. As the wires begin to ablate, the current in the wires permeates into the lower impedance coronal plasma surrounding the wire cores and interacts with the large cavity field, producing a $\mathbf{J} \times \mathbf{B}$ force that accelerates the plasma radially outwards. This causes a subsequent reconfiguration of the magnetic field topology, with the closed loop flux ejected from the wire position by the leading blob of plasma, which is then followed by a steady stream of plasma carrying a radially extended current distribution and entirely open/global field lines embedded in the flow [18]. Within a short distance of the array, the individual streams from the wires merge into a collective flow, reducing azimuthal density modulations arising from these discrete plasma sources [19]. Thus, a cylindrically diverging flow of magnetized plasma is established, which propagates into the surrounding field-free region carrying magnetic field lines which are perpendicular to the direction of the flow [figure 1(b)].

---

* "Inverse" with respect to the operation of an imploding, "Z-pinch" wire array [58,59], as the inverse array creates an expanding plasma flow.

*2.2 Measurements of the properties of the flow*

To demonstrate the presence of the advected magnetic field in the flow and verify that it remains well frozen into the flow on experimental timescales (i.e. several 100 ns) measurements have been made by placing inductive ("B-dot") probes in the path of the flow at several radial distances from the wires [19]. The probes were formed of bound pairs of identical, concentric loops, with an area of 2.7 mm$^2$ (1.55 × 1.75 mm), which were placed parallel to the flow direction (i.e. linking the expected azimuthal direction of the flux). The loops were wound oppositely to provide opposite polarities for the magnetic component of the pick-up, and the raw (unintegrated) signals were recorded separately to visually inspect that these magnetic components were clearly discernible from any possible (noninverted) electrostatic pick-up due to variations of the plasma potential. The probes were shielded by a thin layer of conductive paint to suppress the electrostatic pick-up, which was removed by subtracting the separate loop signals:

$$V_{\text{pair}} = \frac{V_1 - V_2}{2} = \frac{\left(\iint_{A_1} \frac{d\mathbf{B}}{dt} \cdot d\mathbf{A} + \oint \mathbf{E}_{\text{plasma}} \cdot d\mathbf{l}\right) - \left(-\iint_{A_2} \frac{d\mathbf{B}}{dt} \cdot d\mathbf{A} + \oint \mathbf{E}_{\text{plasma}} \cdot d\mathbf{l}\right)}{2} = \frac{d\Phi}{dt}, \quad (1)$$

where A is the loop area, E$_{\text{plasma}}$ is the electric field in the plasma and $\Phi$ is the magnetic flux linked to the probe.

Figure 1(c) shows the combined voltage (blue trace) from a probe pair placed at a radial distance of 8 mm from an aluminium wire array (dimensions: 15 mm length, 20 mm diameter, 8mm diameter cathode, 40 μm diameter wires). At around 30 ns after the start of the current applied to the array, the probe shows a short burst of signal, believed to be from the pick-up of an electromagnetic pulse (EMP) caused by the sharp drop in resistivity of the array circuit as the wire surfaces undergo their rapid phase transition and the current switches into the plasma. A further 50 ns later the voltage measured by the probes begins to rise sharply. This time interval agrees well with the time of flight required for the first plasma to reach the radial position of the probe at flow velocity of V=160 km/s, which is comparable to independent measurements of the flow velocity with Thomson scattering [19,20] and time-gated, multi-frame optical imaging of the position of plasma front [21].

The voltage pulse measured by the inductive probe is proportional to dB/dt of the flux linked to the probe loop and closely replicates the shape of the dI/dt signal of the current through the return electrodes of the wire array [red trace in figure 1(c)], measured using a pair of Rogowski coils. This indicates that the driving current is the source of the magnetic field carried by the flow. Measurements from probes placed at greater radial distances (up to 25 mm from the wires) show increased time offsets before the start of the main signal and maintain a comparable pulse shape, indicating that the field remains frozen to the flow over this distance. To reveal the time-evolution and estimate the strength of the advected magnetic field, the probe signals were numerically integrated yielding field strengths in the range of 2 – 4 T over the experimental timescale. Whilst it is possible that the field strength in the flow is affected by the perturbative nature of this probe measurement, it is worth noting that these values are consistent with independent estimates of the magnetic field strength using a Faraday rotation diagnostic which does not disturb the flow [22].

Further characterisation of the plasma flow from the inverse wire array setup has been made using a suite of spatially and temporally resolved laser probing diagnostics [19,22,23]. These include Mach-Zehnder interferometry imaging for line-integrated measurements of the free electron density, and optical Thomson scattering to obtain localized measurements of the flow velocity and plasma temperatures. Table 1 summarizes the measured properties of the plasma flow for three wire materials commonly used in our experiments. Key attributes are that the plasma flow is in all cases supersonic, internally collisional (mean free path << length scale of the flow) and has a large magnetic Reynolds number (supporting the frozen-in flux condition). The variation in the flow properties with wire material highlights the ability of the setup to tune the flow parameters to different types of experiment and investigate the effects that these properties have on various flow interactions. In the next section we give examples of experiments looking at different interactions of these magnetized flows.

**Table 1.** Parameters of the plasma flow from inverse arrays of carbon (C), aluminium (Al) and tungsten (W) wires [13,19,20,24–27]. Ranges indicate variability of the flow parameters over radial distance from the wires (~5-15 mm) and over the experimental timescale (~100-400 ns). Values quoted for carbon are applicable to magnetic reconnection experiments discussed in section 3.1.

|  |  | Material | | |
| --- | --- | --- | --- | --- |
| Parameter | Symbol | C | Al | W |
| Electron density (cm$^{-3}$) | $n_e$ | 2-5×10$^{17}$ | 0.3-2×10$^{18}$ | 0.3-2×10$^{18}$ |
| Magnetic field (T) | B | 3 | 2-4 | 2-4 |

| Flow velocity (km/s) | V | 50 | 50-150 | 50-150 |
|---|---|---|---|---|
| Ion temperature (eV) | $T_i$ | 50 | 20 | - |
| Electron temperature (eV) | $T_e$ | 15 | 15 | 6 |
| Ionization | $\bar{Z}$ | 4 | 3.5 | <6 |
| Magnetic Reynolds number | $Re_M$ | 20 | 50 | 10 |
| Alfvén speed (km/s) | $V_A$ | 70 | 30 | 10 |
| Ion sound speed (km/s) | $c_S$ | 30 | 20 | 5 |
| Dynamic Beta | $\beta_{dyn}$ | 1 | 10-100 | 200 |
| Thermal Beta | $\beta_{th}$ | 0.4 | 1 | 0.5 |
| Ion-ion mean free path (mm) | $\lambda_{ii}$ | $10^{-2}$ | $10^{-3}$ | $10^{-5}$ |
| Radiative cooling time (ns) | | 100 | 20 | <1 |

## 3. Magnetized plasma flow interaction experiments

### 3.1 Magnetic reconnection in counter-streaming magnetized plasma flows

Magnetic reconnection is a complex and multifaceted process which is an active topic of research across space, astrophysics and plasma physics [28,29]. One of the most notable applications of the inverse wire array setup therefore has been to create a platform for studying this process in the laboratory [25,30–33]. In these experiments two identical inverse wire arrays are fielded side-by-side on the MAGPIE generator, connected in parallel and in the same polarity to each receive a 50% share of the current supply. The two cylindrically expanding plasma flows from these arrays collide in the mid-plane with their respective azimuthal magnetic fields directed oppositely [figure 2(a)]. This produces an extended reconnection layer where the frozen-in flux condition of the advective, upstream flows breaks down, allowing a release of the stored magnetic energy and the restructuring of the magnetic field topology.

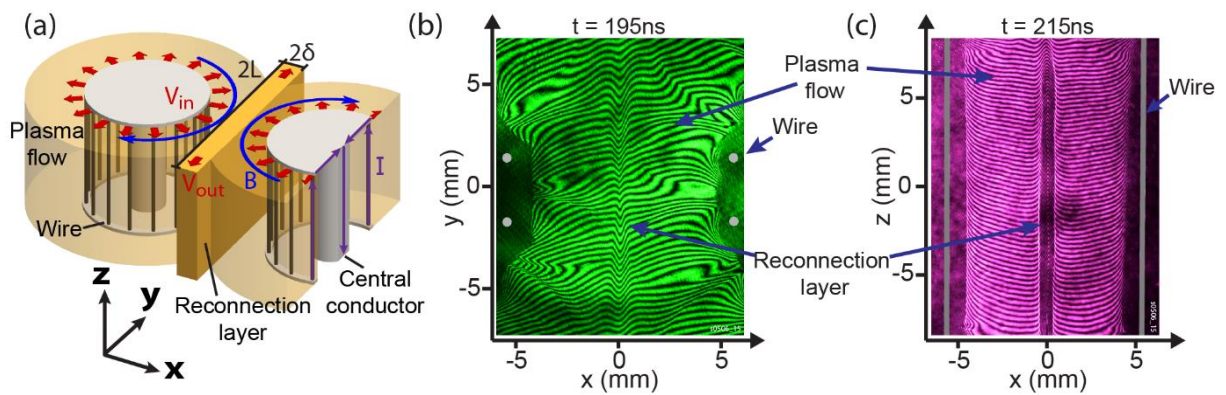

**Figure 2.** (a) Schematic diagram of magnetic reconnection between two wire arrays (with cut-away of the right wire array): current is applied in parallel to two inverse wire arrays, producing magnetized plasma flows which collide to create a reconnection layer. The directions of the current (purple), plasma flows (red), and the embedded magnetic fields (blue) are shown. [(b) and (c)] Raw interferometry

images of the interaction region following the formation of the reconnection layer, showing (b) the xy-plane and (c) the xz-plane, as defined by the Cartesian coordinate system in (a). The positions of the wires are indicated.

The wire array setup is well suited to making experimental observations and measurements of magnetic reconnection as the interaction of the flows occurs in a simple geometry; the layer is produced by symmetric, counter-streaming plasma inflows that are uniform along the axial length of the arrays, and the reconnection plane (in which the reconnecting field lines lie) is well-defined without complications of intricate 3D structure. This is ideal for the accompanying laser probing diagnostics [22], as they make path integrated measurements through the plasma, which can easily be interpreted to understand the structure of the interaction and extract the plasma parameters of interest. In demonstration of this figures 2(b) and 2(c) show interferograms of the colliding plasma flows between two aluminium wire arrays [30,32]. Figure 2(b) was obtained by probing vertically along the axial length of the interaction (z-direction) and figure 2(c) was obtained by probing horizontally (along the y-direction) through the same region. The displacement of the initially straight, horizontal fringes in these interferograms is proportional to the integrated electron density along the probe path [34]. Both images indicate an enhanced electron density inside the layer with respect to the upstream flows, which is uniform along the layer in both orthogonal directions. The cross-sectional density distribution of the reconnection (xy) plane can therefore easily be calculated by dividing the integrated density obtained from figure 2(b) by the axial length of the array [30,32].

Localized measurements of the flow velocities and electron and ion temperatures come from the optical Thomson scattering diagnostic [22]. Together with the interferometry density measurements and Faraday rotation measurements of the magnetic field distribution, the energy exchanges in the system have been analysed [25,30–33]. As the interaction of the magnetized flows occurs over a relatively long timescale, with a continuous inflow of the magnetized plasma for many times the hydrodynamic timescale of the system ($T \gg L/V_{flow}$), the power balance can be monitored as the reconnection process develops and the system evolves. The results have shown that there is a transfer of the kinetic and magnetic energy of the inflowing plasma to the kinetic and thermal energy inside the layer [25,30–32].

The directed kinetic energy of the layer is manifested by fast, symmetric outflows from the centre of the layer, which reach a velocity greater than the upstream Alfvén velocity. This is in accordance with the generalised Sweet-Parker model of [35,36], where the thermal pressure gradient due to the expansion of the plasma into a vacuum provides an additional acceleration to the magnetic tension force driving the outflows. Inside the layer the ions undergo strong anomalous heating, such that the plasma reaches an energy partition of $T_i \sim \bar{Z} T_e$ (≈300-600 eV) (where $\bar{Z}$ is the average ionization). The mechanism for this heating is however not currently understood, as the measured ion temperature is much greater than can be accounted for by shock stagnation of the flows or classical viscous ion heating due to the high velocity shear between the layer and upstream plasma [31,32]. The heating is therefore expected to draw directly from the released magnetic energy in the reconnection current sheet. This however would require a resistivity approximately 10× greater than the Spitzer resistivity to provide heating on a fast-enough timescale to heat the plasma before it leaves the reconnection layer [31,32]. Anomalous heating has been observed in numerous other reconnection experiments (see review of [28] and references therein), and could potentially arise from the development of instabilities in the reconnection layer.

The current drift velocity inferred from Faraday rotation measurements of the magnetic field profile of the reconnection layer:

$$U_d = \frac{J}{en_e} = \frac{1}{e\mu_0 n_e}\frac{\partial B_y}{\partial x},$$

is greater than the local sound speed of the plasma. This could lead to the development of kinetic turbulence (e.g. ion-acoustic or lower hybrid drift instabilities [37]) inside the reconnection layer which could operate as a heating mechanism. Further experiments are in progress to directly measure the drift current velocity using Thomson scattering, as well as searching for signs of the enhanced scattering signal expected with kinetic turbulence effects [38,39]. The preferential heating of ions could be related to the presence of large ion velocities in the direction of the reconnection electric field [$E_z = J_z/\sigma$, positive z-direction in figure 2(a)], as measurements from Thomson scattering in this out-of-plane direction indicate that the ions move with a sufficiently large velocity to carry a substantial fraction of the current inferred from the Faraday rotation measurements [32].

Another important feature in the investigative capability of this reconnection platform is the ability to control the properties of the plasma inflows through the choice of the wire array material and to observe the effects this has on the reconnection layer behaviour. One characteristic difference between the flows generated by carbon and aluminium wire arrays is the ratio of the kinetic (ram) and magnetic pressures of the flows, as defined by the dynamic plasma Beta parameter:

$$\beta_{dyn} = 2\mu_0 \rho V_{flow}^2 / B^2. \qquad (3)$$

In carbon wire experiments these pressures are almost equal ($\beta_{dyn} \sim 1$), whereas for aluminium the collision of the plasma flows is more strongly driven by the kinetic pressure ($\beta_{dyn} \sim 10$), which has the effect of creating a pile-up of magnetic flux at the boundary of the reconnection layer [32]. The most substantial difference between these two cases, however, is the susceptibility of the reconnection layer to the plasmoid instability [40]. A key parameter governing the stability is the Lundquist number, which is defined as the ratio of the Alfvén crossing timescale of the system to the resistive timescale:

$$S = LV_A/\eta, \qquad (4)$$

where L is the length scale of the system, $V_A$ is the Alfvén speed and η is the magnetic diffusivity. In the aluminium experiments the radiative cooling of the plasma maintains a relatively low electron temperature of 30 eV inside the reconnection layer. This results in an increased magnetic diffusivity and a modest Lundquist number (S~10), equating to a stable layer [figure 3(a)]. In carbon experiments however, the larger electron temperature of 60 eV results in a Lundquist number of S~100, which is sufficiently large that it is expected to place the layer in the plasmoid unstable, semi-collisional reconnection regime [41,42]. This is consistent with observations of the layer structure, which show the formation of plasmoids in both self-emission images [figure 3(b)] and laser interferometry [25,31,33]. Measurements with B-dot probes placed in the exhaust of the layer have indicated these indeed possess a magnetic O-point structure [25,31].

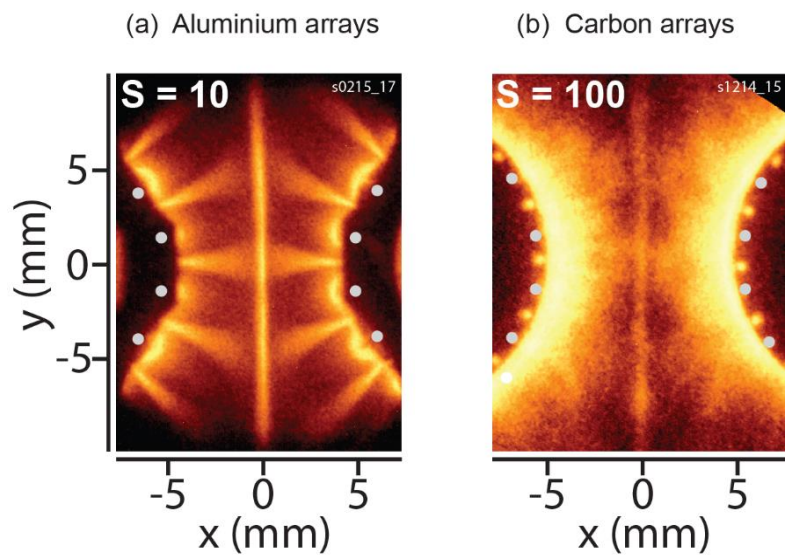

**Figure 3.** Optical self-emission images (false-colour) of the reconnection plane from two experiments with different Lundquist numbers, due to the different flow parameters produced by the selected wire array material. Wire positions are marked in white. (a) Aluminium wire arrays produce a stable reconnection layer of Sweet-Parker type appearance. (b) Carbon wire arrays produce a layer which is unstable to the production of plasmoids. Multi-frame videos showing the dynamics of the respective layers over each experiment can be found in the references [32] and [25].

*3.2 Interactions of the magnetized flow with conducting and dielectric obstacles*

The supersonic and super-Alfvènic velocities which can be achieved by the flow from an inverse wire array (see table 1) make it suitable for creating magnetized shock structures. Experiments were carried out to study magnetized shocks in a simple interaction geometry by placing either a planar or cylindrical obstacle normal to the path of the plasma flow [figure 4(a)] [19,26,43]. In order to ensure that the magnetic field played a role in the structure of this shock, a conducting (metallic) obstacle was used, as this provided a boundary to prevent the magnetic flux diffusing through the obstacle surface, therefore forcing the flux to accumulate with the post-shock plasma.

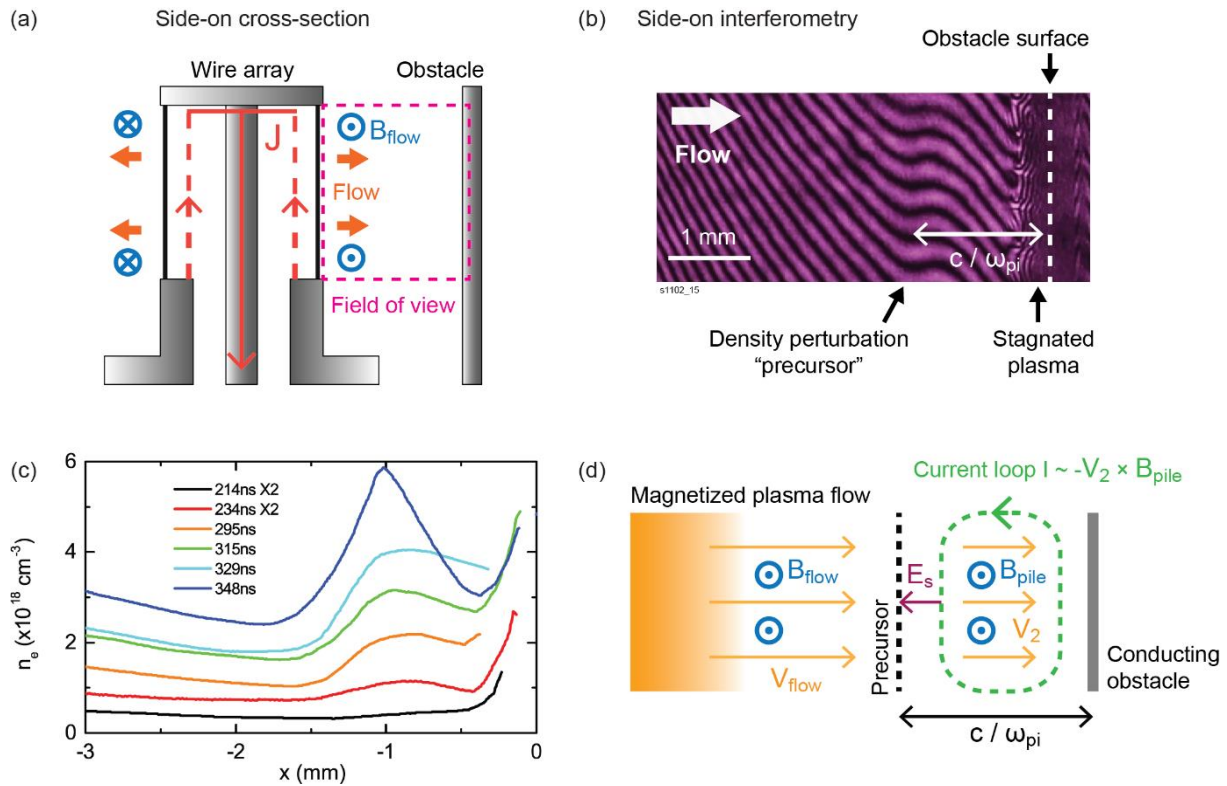

**Figure 4.** (a) Schematic diagram for the setup of magnetized shock experiments with conducting obstacles. An extended metallic obstacle is placed in the path of the plasma flow from an inverse wire array, creating a perpendicular magnetized shock with uniformity in the axial direction. (b) Side-on interferogram showing the key features of the shock structure. (c) Time series of the electron density profile along the flow direction, demonstrating the steeping of the precursor transition. (d) Schematic of the development of the precursor by the pile-up of the advected magnetic field at the obstacle surface.

Figure 4(b) shows an interferogram of the interaction of the flow from an aluminium wire array with a copper obstacle (placed at 10 mm from the array), which was obtained by probing along the "side-on" line-of-sight indicated in figure 4(a). The predominant features observed in this interaction are a dense layer of stagnated plasma formed by the collision of the plasma at the obstacle surface and a detached electron density perturbation, or "precursor", in the upstream flow. This precursor is observed to form at this position, becoming detectable approximately 50 ns after the stagnation of the flow [18]. Thereafter it maintains this fixed distance from the obstacle, whilst gradually steeping in density profile; transforming from a small density increase into a larger step-like transition consistent with that of a fully developed shock front [figure 4(c)].

Measurements with both inductive ("B-dot") probes and Faraday rotation show a pile-up of the magnetic field in the region between the precursor and the obstacle surface [19,43]. This accumulated field is ultimately responsible for holding off the ram pressure of the plasma flow ($B^2/2\mu_0 \approx \rho V_{flow}^2$). However, the magnetic field is not fully localized in the stagnated plasma but is able to resistively diffuse into the oncoming flow. A calculation of the gyro-radii for the electrons and ions in the incoming flow shows that the field is strong enough to directly localize the electrons ($r_{g,e} = V_{th}/\omega_{ce} \approx 5 - 10\ \mu m \ll x_s$), but not the ions ($r_{g,i} = V_{flow}/\omega_{ci} \approx 5 - 10\ mm \gg x_s$): where $\omega_c$ are the cyclotron frequencies and $V_{th}$ is the electron thermal velocity and $x_s$ is the stand-off distance of the precursor from the obstacle surface [44].

The observed stand-off distance is comparable to the ion inertial length of the plasma ($d_i = c/\omega_{pi} \approx 2.5\ mm$, where $\omega_{pi}$ is the ion plasma frequency). This is the scale at which ions in the plasma can decouple from the electrons, such that the magnetic field remains frozen into only the electron fluid rather than the bulk of the plasma. We therefore interpret that the precursor is formed by the build-up of magnetic field in this region, which acts on the incoming electrons, causing them to decelerate whilst the unmagnetized ions pass through initially unimpeded. Over time this growing two-fluid separation is expected to produce an electric field, which in turn decelerates the flow of ions causing the transition into a shock front separate from the main stagnation shock [figure 4(d)]. It is noted that these experimental observations are pertinent in the context of space physics as they provide support for a fast-flow braking mechanism previously postulated to occur in auroral magnetospheric events [43,45], and also bear some similarity to solar wind interactions with planetary ionospheres, which act as obstacles with highly conducting boundaries, leading to induced magnetospheres around planets and moons which do not possess magnetic cores [46,47].

To further demonstrate the fundamental importance of magnetic flux pile-up in mediating the precursor development, experiments were also performed to provide a comparison of the shock structures produced when the flow impacts conducting and dielectric (insulating) obstacles of equivalent size and geometry [26]. Figure 5 shows images from an experiment with 500 μm-diameter copper and glass cylindrical obstacles, placed 7 mm apart in the flow from a single (aluminium) wire array. The interaction is viewed from the "end-on" perspective, looking down along the axes of the obstacles, which were parallel to the axis of the wire array, such that the magnetic field lines of the flow were aligned perpendicular to the obstacle axes. The bow shocks formed around the two obstacles show significant differences in their shape and stand-off distances. The bow shock formed around the glass obstacle is much closer to the front surface of the cylinder (< 100 µm), and possesses a narrow opening angle of 22°, which is consistent with the Mach angle for a hydrodynamic shock at the measured sonic velocity of $M_S$=2.5. This indicates that the resistivity of the plasma at the obstacle surface was high enough to allow the magnetic flux to freely diffuse through the obstacle and therefore not affect the shock structure. In comparison the bow shock around the conducting obstacle stands-off at the ion inertial length of ~1.5 mm. It has a much wider opening angle (45°), indicating that the bow shock is supported by the tension force of the magnetic field lines which are draped around the obstacle by the flow, and cannot escape into the downstream region. Similar draping effects are detected around planets, moons, solar coronal mass ejections, and galaxies [46–50].

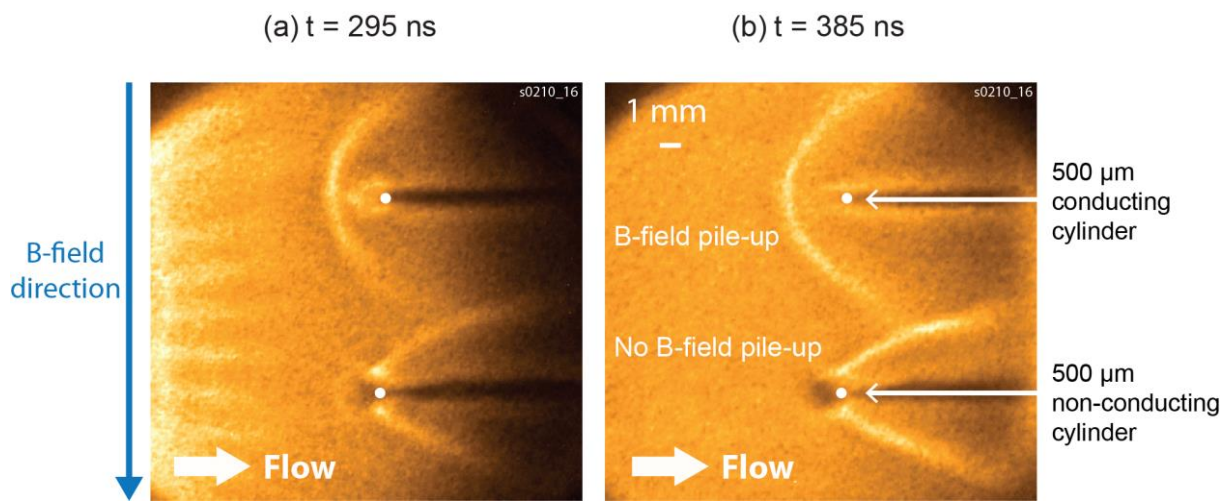

**Figure 5.** Optical emission images (false-colour) showing the contrasting bow shocks formed around conducting and dielectric cylindrical obstacles (obstacle axes into page). The shocks are formed side-by-side in the same experiment. The images at different times during the experiment indicate that the bow-shock structures are dynamically stable.

*3.3 Interactions of the magnetized flow with magnetized obstacles*

In this section we present results from experiments extending the investigation of flow-obstacle interactions to a configuration where both the plasma flow and the obstacle have their own magnetic fields [51]. This type of interaction is relevant to the study of the interplay between stellar winds and the magnetospheres of planetary bodies [2,52–55] or exoplanets [56].

To produce strongly magnetized obstacles in these experiments, we employ a fraction of the current from the pulsed-power generator. The inverse wire array and the obstacle are connected to the current supply in parallel, and the current split between these is determined by the relative inductances of their current paths. The inductances are selected to provide a dynamically significant magnetic field pressure around the obstacle, without reducing the current through the wire array enough to significantly alter the properties of the plasma flow in comparison with previous experiments (i.e. the flow properties remain similar to table 1). Figures 6(a) and 6(b) show cross-sectional schematic diagrams of the setup from the first set of experiments to demonstrate this principle. The obstacle is a cylindrical stainless-steel post of 5 mm diameter, set at an axis-to-axis distance of 22.5 mm (10 mm wire to obstacle surface) from a 20 mm diameter array of 16 × 40 μm aluminium wires. The vertical height of the interaction region (i.e. axial length of the array and obstacle) is 20 mm. The current was applied to the cylinder such that the azimuthal magnetic fields of the plasma flow and the obstacle were parallel in the interaction region (as depicted in figure 6). The versatility of the current supply however allows future experiments to operate with reverse obstacle B-field polarity to investigate the interaction with opposing field directions. The current through the obstacle and the total current supplied were each monitored by pairs of Rogowski coils on the return paths to ground. The current through the obstacle was measured to be approximately 25% of the total current (corresponding to a field strength of 15 T at the surface of the obstacle at peak current). Null experiments were also run without any current supplied to the obstacle, to provide a geometrically identical comparison to the shock structure with the unmagnetized, conducting obstacle (as in section 3.2).

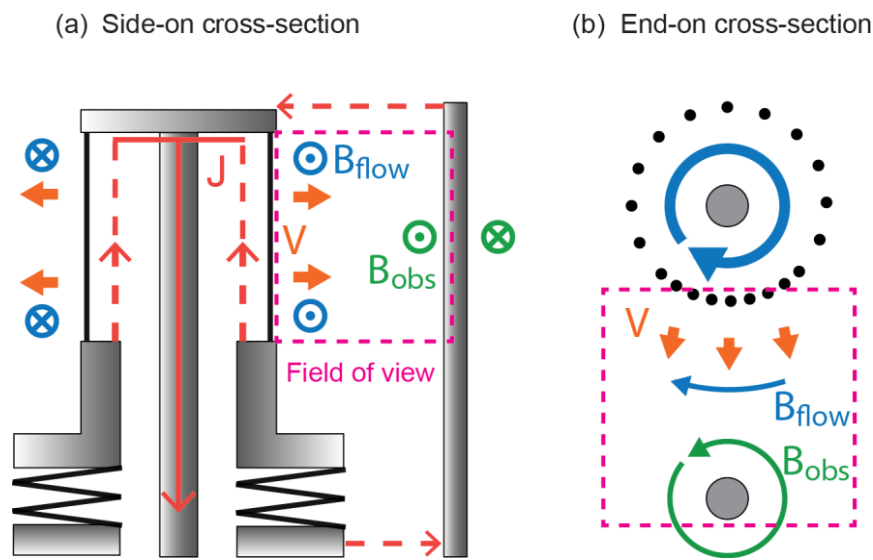

**Figure 6.** Schematic diagrams for the setup of magnetized shock experiments with a cylindrical obstacle that possesses its own magnetic field. A fraction of the drive current is supplied in parallel through the obstacle to produce the private magnetic field. The (a) side on, and (b) end on, cross-sections indicate the fields of view for the orthogonal laser probing diagnostics.

Figures 7(a) and 7(b) show side-on interferograms of the interaction region indicated in figure 6(a) at around peak current (t=250 ns) for the null and magnetized obstacle experiments respectively. It is apparent that the stand-off distances of the magnetic precursors formed ahead of the obstacles in these two experiments differs by approximately a factor of two. In the image of figure 7(a) the precursor stands off from the unmagnetized, conducting obstacle at a distance of 2.2 mm ≈ $c/\omega_{pi}$. This is consistent with the mechanism described in section 3.2, whereby the precursor is supported by the pile-up of the advected magnetic flux and the resulting two-fluid separation. In the case of the magnetized obstacle in figure 7(b) however, a precursor appears in the flow at almost twice the stand-off distance of 4 mm from the obstacle surface. This demonstrates that the strong magnetic field of the obstacle is able to influence the flow much further upstream. This is best seen by observing the interaction from the "end-on" perspective, looking down onto the region identified in figure 6(b), as this makes use of the uniformity of the setup along the axial direction. Data obtained from this perspective are presented in figure 8.

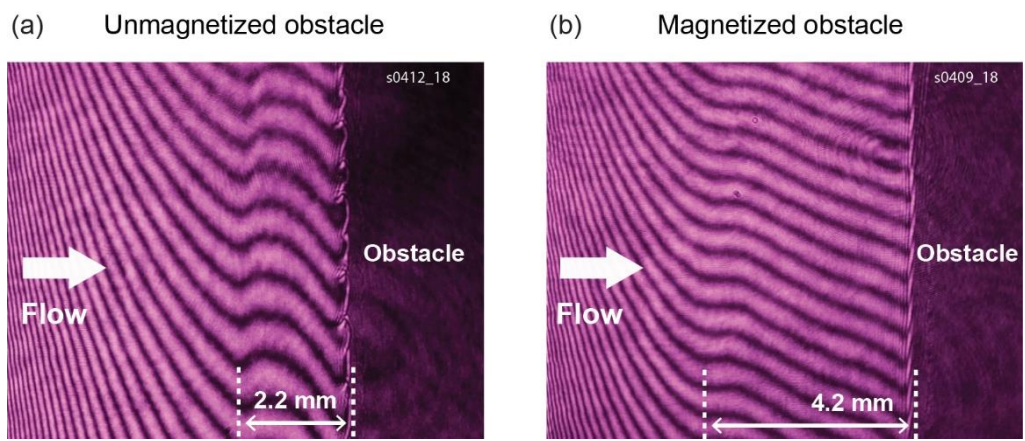

**Figure 7.** Side-on interferograms of the interaction of the magnetized flow with (a) unmagnetized, and (b) magnetized cylindrical obstacles. The stand-off distances of the precursors are indicated.

Figures 8(a) and 8(b) show end-on images of the optical self-emission of the plasma for the same two experimental cases in figure 7 (and at similar times in the experiments). The images were captured using a high-speed, multi-frame, optical camera which allows the qualitative features of the shock structure to be observed. Videos are also available in the online supplementary material, which show the formation and evolution of the shock structure over the full observed time range of t=90-420 ns (Δt=30 ns) from each of these experiments. In the field of view the flow of magnetized plasma is moving down from the wire array at the top edge of the images towards the obstacle indicated at the bottom of the images. The flow closest to the wire array displays repeating wedge-shaped features arising from the oblique shocks between the streams from neighbouring wires, which merge into a collective flow as the plasma moves downstream. An emitting layer of plasma is visible at the front surface of the obstacle in both images, which is most likely due to the stagnation of the flow, with some further possible contribution from radiative heating of the obstacle surface by XUV emission from the wire array. The comparable thickness and brightness of this layer in the two images however indicates that this surface plasma does not arise from resistive heating of the obstacle by the applied current.

The most significant difference observed between the two experiments is indeed in the position of the aforementioned precursors. The end-on perspective images reveal the cross-sectional shape of these precursors. In the image of figure 8(a) the conducting obstacle without the applied magnetic field displays a similar wide opening angle bow shock precursor as that observed for the conducting cylindrical obstacle of smaller diameter in figure 5; and so we attribute this to the draping of the magnetic field lines around the obstacle as discussed in section 3.2. In the case of the magnetized obstacle in figure 8(b) the precursor is harder to make out as its position further upstream overlaps with structure of the merging wire streams; however, it appears to display a similar curvature.

To provide quantitative measurements of the plasma structure laser interferometry was used to probe along the same line of sight. Figures 8(c) and 8(d) show the raw interferograms from the same experiments and for the same field of view as figures 8(a) and 8(b), at the slightly earlier time of t=210 ns after current start. To plot the distribution of the free electron density the interferograms were processed using the technique described in references [22,23,57]. Figures 8(e) and 8(f) show the resulting maps of the electron density. Comparison of the two images shows a clear build-up of plasma at large distance from the magnetized obstacle, with flow densities that are considerably higher than with the unmagnetized obstacle over the range of y=4-8 mm from the obstacle centre. Behind this build-up, downstream of the bow shock (y<4 mm), the magnetized obstacle creates a relative void of plasma material. The position of the magnetopause agrees well with the position at which the ram pressure of the flow (characterised in past experiments [6]) matches the calculated magnetic pressure of the cylinder (using Ampère's law and the measured current), suggesting that the magnetic field holds off the flow. This structure is reminiscent of the abrupt boundary of the magnetopause between a planetary magnetosphere and the plasma flow of the solar wind, which is redirected around the planetary body by this magnetic field [2].

| Unmagnetized obstacle | Magnetized obstacle |
|---|---|
| 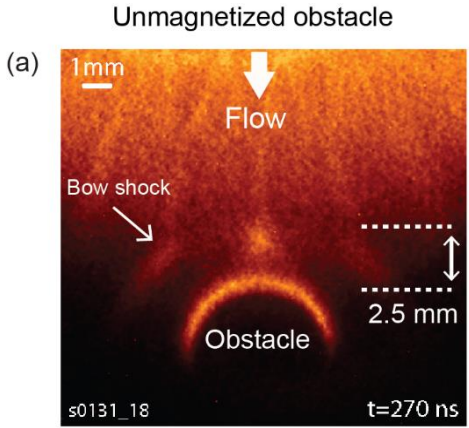 | 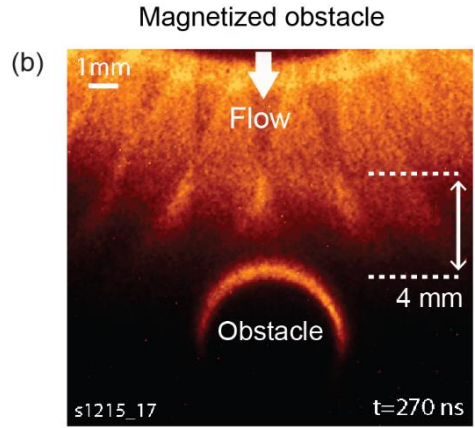 |
| 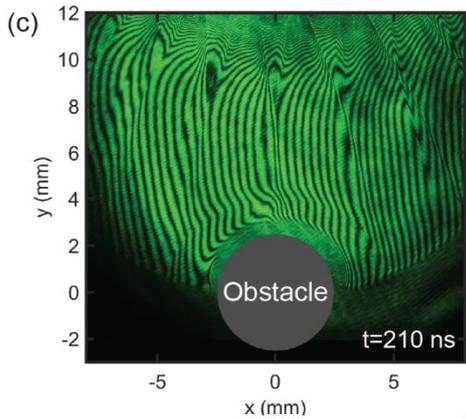 | 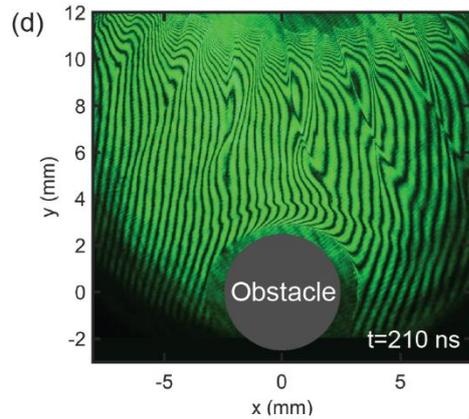 |
| 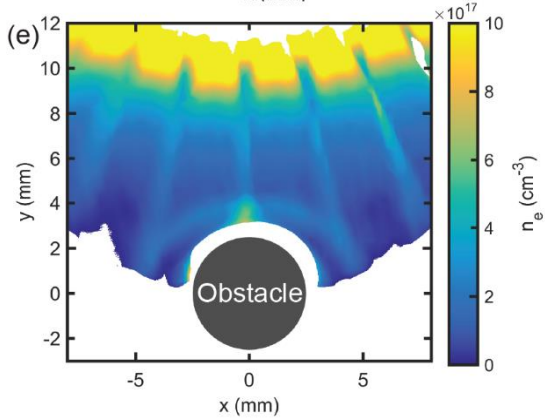 | 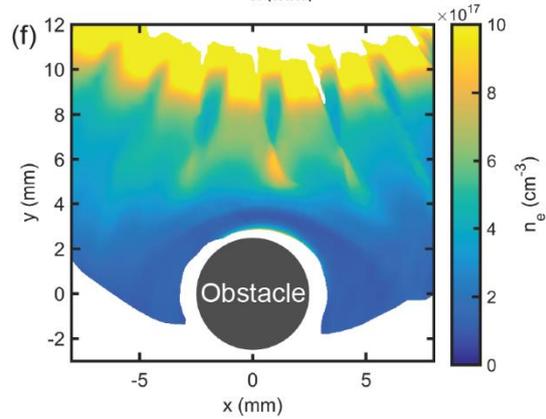 |
| 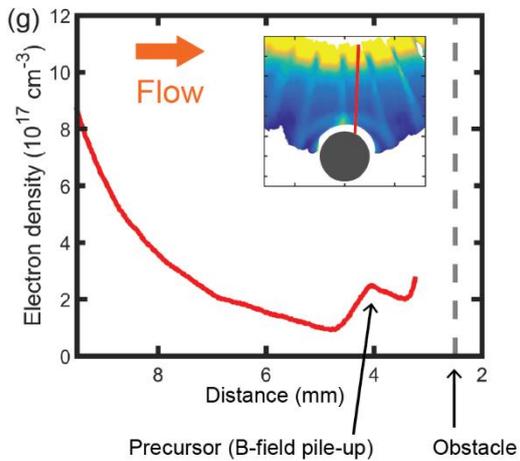 | 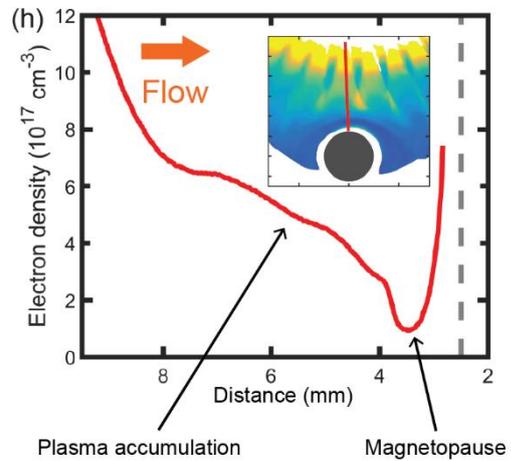 |

**Figure 8.** Comparison of the structure of shocks formed around unmagnetized (conducting) and magnetized obstacles, showing the field of view identified in figure 6(b). (a & b) Optical self-emission images showing differences in the bow-shock stand-off distances. (c & d) Raw interferograms. (e & f) Processed electron density maps. (g & h) Electron density profiles of the flow along the chords identified. Key features of the interactions are indicated.

The difference in the spatial profiles of the shock structures from the two experiments is best seen in figures 8(g) and 8(h), which present lineouts of the electron density along the chords marked in figures 8(e) and 8(f). Clearly the obstacle magnetic field alters the shock structure over an extended range, redistributing the plasma density to distances much further from the obstacle body. The next step in this investigation will be to acquire further quantitative measurements of the structure by employing Thomson scattering across this range of the upstream flow to measure the effect of the magnetic field on the flow velocity. Thomson scattering will also allow us to compare plasma temperature distributions in the two structures.

Looking further ahead we aim to perform experiments with different and more complex interaction geometries and magnetic field orientations. This could include obstacles with dipolar magnetic fields, more relevant to the three-dimensional magnetized interaction structures observed in space. Interactions of the magnetized flows with gaseous or plasma states are also possible, to replicate interactions with interstellar medium and gas/dust clouds.

## 4. Summary

In this paper we have provided a detailed description of a versatile pulsed-power driven setup for creating supersonic flows of magnetized plasma and given examples of high energy density physics experiments on the MAGPIE generator studying the interactions of magnetized plasma flows. The inverse wire array setup consists of a cylindrical arrangement of thin wires connected between two electrodes. An applied current of 1 MA converts the wires to a plasma, which is accelerated radially outward by the $\mathbf{J} \times \mathbf{B}$ force, carrying a frozen-in azimuthal magnetic field of several Tesla. As the plasma flow expands it propagates into an initially field-free region, which is highly accessible for diagnostic measurements of the plasma properties. The flow is steady and continuous over the duration of the applied current pulse (∼500 ns), allowing the study of the long-term behaviour of the flow.

The first example given of experiments performed with inverse wire arrays was that of magnetic reconnection in colliding plasma flows. These experiments make use of two side-by-side wire arrays to produce counter-streaming flows with oppositely directed magnetic fields. The interaction of the flows leads to the formation of a magnetic reconnection layer, in which ions are heated by the release of the stored magnetic energy through an anomalous mechanism. The structure of the reconnection layer is highly dependent on the properties of the inflowing plasma. The inflow properties can be varied by the choice of wire array material, which alters the Lundquist number of the system, allowing access to regimes where the layer is either stable or unstable to the formation of plasmoids.

Magnetized shock experiments have also been carried out by placing obstacles in the path of the plasma flow. When a conducting obstacle is used this leads to a pile-up of the magnetic flux at the obstacle surface, extending ahead of the stagnated plasma. As this magnetic field accumulates it acts on the magnetized electrons of the incoming flow, while the ions remain unmagnetized. This two-fluid separation creates a growing electric field, decelerating the ions and causing a density transition to develop at $c/\omega_{pi}$ from the obstacle that gradually steepens into an abrupt shock front. Experiments with non-conducting, dielectric materials do not exhibit shock fronts with these large stand-off distances and their bow shocks show much narrower opening angles, due to the lack of magnetic field pile-up and draping of the field over the obstacle.

The latest inverse wire array experiments have extended the investigation of magnetized shocks to study the interactions of the magnetized flow with obstacles possessing their own magnetic fields. The obstacles are subject to an applied current, which produces a magnetic field geometry determined by the current path through the obstacle. Initial experiments have looked at the structure of the shock formed when the flow interacts with a cylindrical obstacle, whose azimuthal magnetic field is parallel to the field embedded in the flow. This field was observed to affect the plasma flow much further upstream than the pile-up field accumulated with an unmagnetized, conducting obstacle. Additionally, the magnetic pressure around the obstacle created a void of plasma behind the bow shock, analogous to the magnetopause around a magnetized planetary body. Future experiments will investigate the effect of the orientation and geometry of the obstacle field on the structure of the interaction and make quantitative measurements of the motion of the plasma around the obstacle body using Thomson scattering.

**ACKNOWLEDGEMENTS**

This work was supported by the U.S. Department of Energy (DOE) Award Nos. DE-NA0003764, DE-F03-02NA00057 and DE-SC-0001063, and by the Engineering and Physical Sciences Research Council (EPSRC) Grant No. EP/ N013379/1. NFL was supported by the NSF-DOE partnership in basic plasma science and engineering, award no. DE-SC0016215.